\documentclass[%
reprint,
amsmath,amssymb,
aps,
prb,
]{revtex4-2}

\usepackage{graphicx}
\usepackage{dcolumn}
\usepackage[force]{feynmp-auto}
\usepackage{multirow}
\usepackage{color}

\usepackage[margin=0.6in]{geometry}

\usepackage[caption=true]{subfig}
\usepackage{float}

\newcommand{\ie}{{\it i.e., }}

\newcommand{\first}{1$^{\text{st}}$ }

\begin{document}
	
	\preprint{APS/123-QED}
	
	\title{Beyond the Tamura model of phonon-isotope scattering}
	
	\author{Nakib H. Protik}
	\email{nakib.haider.protik@gmail.com}
	\affiliation{Department of Physics and CSMB, Humboldt-Universität zu Berlin, 12489 Berlin, Germany}
	\author{Claudia Draxl}%
	\email{claudia.draxl@physik.hu-berlin.de}
	\affiliation{Department of Physics and CSMB, Humboldt-Universität zu Berlin, 12489 Berlin, Germany}
	\affiliation{Max Planck Institute for Solid State Research, Heisenbergstraße 1, 70569 Stuttgart, Germany}
	
	\date{\today}
	
	\begin{abstract}
		The Tamura model is a particular type of \first Born approximation of the phonon-isotope scattering problem. The expression for the mode-resolved phonon-isotope scattering rates in this model, derived in 1983, is still widely used in \textit{ab initio} transport calculations. While the original work emphasized its applicability to low-energy acoustic phonons only, it has, nevertheless, been also applied to optical phonons in the field of phonon transport. The model has the salient feature of being a perturbation theory on top of a virtual crystal background. As such, this approach does not correspond to the proper methodology for solving the phonon-substitution defect problem in the respective limit. Here, we explore three avenues to go beyond the Tamura model and carry out calculations on a set of common materials to compare the different approaches. This work allows a systematic improvement of the treatment of phonon-isotope scattering in \textit{ab initio} phonon transport, while unifying it with the general phonon-substitution scattering problem.
	\end{abstract}
	
	\maketitle
	
	\section{Introduction}
	The presence of an isotopic mixture of elements in a crystal leads to observable changes in its transport properties. A famous historical example can be picked from the field of superconductivity. There, the observation of the isotope effect eventually inspired the formulation of the theory of phonon-mediated superconductivity. In the field of phonon transport, on the other hand, the lattice thermal conductivity can be strongly reduced in a crystal with a natural isotopic mixture of atoms compared to one that is isotopically pure \cite{lindsay2013ab}. Now, in state-of-the-art \textit{ab initio} phonon transport research, the Tamura model \cite{tamura_isotope_1983} is, fourty years after its introduction, still the \textit{de facto} choice for treating the phonon-isotope scattering problem. It is the \first Born approximation of the phonon-isotope scattering problem with the additional requirement that the phonons be calculated using an isotopic average of each atomic species. Every isotope is then treated as a substitutional defect on top of the virtual atom. As such, the phonons are that of a virtual crystal, and the method relies on the virtual-crystal approximation (VCA). 
	
	In recent years, a significant amount of effort has been dedicated to treating the phonon-substitution defect-scattering problem using \textit{ab initio} methodologies \cite{nebil_2014, katre2017exceptionally, protik2016ab, thebaud2020success, polanco2018ab, fava2021dopants}. In general, a substitutional defect leads to an on-site mass-perturbation and an extended-range bond perturbation. The latter is due to the fact that the chemical environment surrounding the substitution atom is different from that around the host atom. One can readily see that the phonon-isotope scattering problem is a special case of the phonon-substitution scattering problem in which the bond perturbation simply vanishes. As such, it is desirable to have a methodology for handling the phonon-substitution scattering problem that is equivalent to the phonon-isotope scattering problem in the appropriate limit. However, the commonly employed Tamura model of the phonon-isotope scattering does not satisfy this correspondence principle.
	
	To illustrate the last statement, we consider the concrete case of Si substitution at the As site in cubic BAs, as discussed in Ref. \cite{fava2021dopants}. In the phonon-substitutional defect scattering methodology, Si is substituted at an As site in the BAs crystal, \ie not at a virtual atom created from an average of Si and As. Similarly, the bond perturbation due to this defect has to be calculated by taking the difference of the $2^{\text{nd}}$ order interatomic force constants between the pristine BAs supercell and one where one of the As atoms has been substituted with a Si atom. Again, no virtual-crystal supercell built from the pristine and defective supercells is ever invoked. In the same material, however, when it comes to treating the phonon-isotope scattering, isotopically averaged, \textit{virtual} B atoms at virtual B sites are created with $19.8\%$ $^{10}$B and $80.2\%$ $^{11}$B. This shows that two divergent approaches are currently being used to treat the same class of scattering problems and demonstrates a failure to satisfy the correspondence principle.
	
	Another inconsistency arises from the fact that the VCA and isotopically pure crystals have different phonon spectra. This is because the VCA, by construction, already includes a phonon renormalization (we will show later that this is an arbitrary choice) due to the isotopic variance. Now, when the effect of the phonon-isotope scattering on the phonon transport is studied, two separate calculations are typically performed -- one with the phonon-isotope scattering included and another with it turned off. However, the same VCA phonon ground-state energies are used in both, although the second calculation should ideally be done on top of an isotopically pure crystal. This practice is usually well justified for materials where the VCA and the isotopically pure system energies are close. Ideally, however, the closeness of the two ground states should always be checked.
	
	In this work, we propose a solution to address these inconsistencies by treating the phonon-isotope scattering as a special case of the phonon-substitutional defect problem. We show how we can go beyond the Tamura model by (i) foregoing the VCA; (ii) going beyond the \first Born approximation of phonon-impurity interaction in the scattering $T$-matrix; and (iii) doing both. We provide an alternative expression to the Tamura formula for the phonon-isotope scattering rates, which can be readily implemented in \textit{ab initio} phonon transport codes. By computing a number of commonly studied materials, we compare the \first and full Born approximations of the scattering $T$-matrix for this problem and show when the \first Born approximation becomes inadequate.
	
	\section{Theory}
	In this section, we present the theory of the phonon-impurity scattering as a special case of the more general phonon-substitution defect scattering problem. We start by describing it for a single point mass-defect at an atomic site in the crystal. Following this, we discuss how to obtain the expression for the scattering rates for a crystal with a collection of such defects. This leads us to a diagrammatic analysis of the impurity-scattering problem and allows us to compare the Tamura model to a certain class of non-perturbative approximations.
	
	\subsection{Scattering from a single defect}
	Following the textbook treatment of scattering \cite{sakurai_modern_1994}, we consider an incident phonon state $|\lambda\rangle$ scattered by a compact potential $V$ to result in a final phonon state $|\lambda'\rangle$,
	\begin{equation}\label{eq:scattstate}
		|\lambda'\rangle = |\lambda\rangle + |\lambda_{S}\rangle,
	\end{equation}
	where $|\lambda_{S}\rangle$ is the scattered part of it. In the case of elastic scattering, the final state is formally related to the incident state via the Lippmann-Schwinger equation,
	\begin{equation} \label{eq:lipp-schwing}
		|\lambda'\rangle = |\lambda\rangle + g^{+}V|\lambda'\rangle,
	\end{equation}
	where $g^{+}$ is the retarded, bare phonon Green's function operator, $V$ is the scattering potential operator, and $|\lambda\rangle \equiv |s\mathbf{q}\rangle$ is the phonon state tagged by band index $s$ and wavevector $\mathbf{q}$.
	
	For a compact scattering potential, which is the case for on-site mass perturbations due to isotopic substitutions, it is convenient for computational reasons to use a real-space representation in the lattice displacement basis. Following Ref. \cite{nebil_2014}, we write 
	\begin{equation} \label{eq:g0}
		g^{+}_{ij}(\omega^{2}) = \sum_{\lambda}\dfrac{\phi^{i}_{\lambda}\phi^{*j}_{\lambda}}{\omega^{2} - \omega_{\lambda}^{2} + i0^{+}} ,
	\end{equation}
	where $\omega$ is a sampling angular frequency and $|i\rangle \equiv |\tau\alpha\rangle$ represents a lattice displacement with $\tau$ identifying the atom in the supercell and $\alpha$ labeling the Cartesian component. $\phi^{i}_{\lambda} \equiv \langle i | \lambda \rangle = \exp(i\mathbf{q}\cdot\mathbf{R}_{i})\xi^{\tau\alpha}_{s}(\mathbf{q})$ with $\mathbf{R}_{i}$ being the position of the unit cell in the supercell that contains the atom $\tau$, and $\xi (\mathbf{q})$ is the phonon eigenvector matrix obtained from diagonalizing the dynamical matrix at wave vector $\mathbf{q}$.
	
	From Eqs. \ref{eq:scattstate} and \ref{eq:lipp-schwing}, we get
	\begin{equation}
		|\lambda_{S}\rangle = g^{+}T|\lambda'\rangle,
	\end{equation}
	where the transition matrix $T$ is given by
	\begin{equation}\label{eq:Tmatrix}
		T = \left( 1 - Vg^{+} \right)^{-1}V.
	\end{equation}
	The expansion of the right-hand side of above equation gives the Born series, which can be interpreted as an infinite sequence of scatterings at the same potential site.
	
	Next, we discuss the potential that results from substituting a host atom with a guest atom. We assume that the chemical environments surrounding the host and the guest are identical and only an on-site mass perturbation is created in the process. This is a reasonable assumption for the case of isotopic substitutions, since isotopes only vary in the number of neutrons. The scattering potential for such a case is given by \cite{tamura_isotope_1983}
	\begin{equation}\label{eq:Vmass}
		V_{\tau\alpha, \tau'\alpha'}(\omega^{2}) = -\omega^{2}\sum_{d_{\tau}}\dfrac{\Delta M^{(d_{\tau})}_{\tau\tau'}}{M_{\tau}}\delta_{\tau\tau'}\delta_{\alpha\alpha'},
	\end{equation}
	where $d_{\tau}$ labels the type of the defect, which in our case are the different isotopes of atom $\tau$, and 
	\begin{equation}\label{eq:massdiff}
		-\dfrac{\Delta M^{(d_{\tau})}_{\tau\tau'}}{M_{\tau}} \equiv -\dfrac{M^{(d_{\tau})}_{\tau} - M_{\tau'}}{M_{\tau}}
	\end{equation}
	is the mass-perturbation matrix due to the substitution. Note that in the Tamura model, the host mass $M_{\tau}$ in the denominator of Eq.~\ref{eq:Vmass} is arbitrarily taken to be the isotope-averaged mass following the VCA. We will discuss this further in Section II C.
	
	\subsection{Scattering from an ensemble of defects}
	The phonon-isotope scattering problem is essentially a problem of describing the scattering of phonons from a collection of 0-dimensional impurities placed at \emph{randomly} chosen atomic sites. We argue that the use of the word ``impurity" strongly implies that most of the system must be pure. In other words, only a minority of the atomic sites can host an isotopic substitution. Now, in order to practically calculate the properties of the defective system, we seek an impurity-averaged Green's function. This is a single-particle Green's function of the defective system which restores the periodicity of the system which was broken by the presence of the impurities. This new Green's function describes a system of impurity-renormalized (quasi)phonons with finite lifetimes. Details of the averaging procedure can be found, for example, in Refs. \cite{mahan2000many} and \cite{bruus2004many}. Here we simply summarize the standard results. 
	
	The impurity-averaged Green's function obeys the Dyson equation
	\begin{equation}\label{eq:dyson}
		\langle g^{+} \rangle_{\text{imp}}^{-1} = \left(g^{+}\right)^{-1} - \Sigma\left[\langle g^{+} \rangle_{\text{imp}}\right],
	\end{equation}
	where $\Sigma$, a functional of the dressed Green's function in general, is the irreducible self-energy of the phonons due to their interaction with the impurities. For a low fractional density of impurities, $f$, the terms linear in $f$ are retained such that
	\begin{equation}\label{eq:lowconcself}
		\Sigma_{\lambda} = T_{\lambda\lambda},
	\end{equation}
	where $T_{\lambda\lambda}$ is the diagonal of the \textit{impurity-averaged} $T$-matrix due to a single defect. In this regime, all interactions at the same impurity site are re-summed. Diagrammatically, this quantity can be represented as
	\begin{equation}\label{eq:T_lowconc}
		T_{\lambda\lambda'} = 
		\begin{fmffile}{line1}
			\parbox{30pt}{
				\begin{fmfgraph}(40,20)
					\fmfleft{i1,i2}
					\fmfright{o1,o2}
					\fmf{phantom}{i1,v1,o1}
					\fmffreeze
					\fmf{phantom}{i2,v2,o2}
					\fmffreeze
					\fmf{dashes}{v1,v2}
					\fmfv{decor.shape=hexacross, decor.size=4thick}{v2}
				\end{fmfgraph}
			}
			+ \parbox{40pt}{
				\begin{fmfgraph}(40,20)
					\fmfleft{i1,i2}
					\fmfright{o1,o2}
					\fmf{plain}{i1,v1,o1}
					\fmffreeze
					\fmf{phantom}{i2,v2,o2}
					\fmffreeze
					\fmf{dashes}{i1,v2}
					\fmf{dashes}{o1,v2}
					\fmfv{decor.shape=hexacross, decor.size=4thick}{v2}
				\end{fmfgraph}
			}
			+ \parbox{40pt}{
				\begin{fmfgraph}(40,20)
					\fmfleft{i1,i2}
					\fmfright{o1,o2}
					\fmf{plain}{i1,v1,o1}
					\fmffreeze
					\fmf{phantom}{i2,v2,o2}
					\fmffreeze
					\fmf{dashes}{i1,v2}
					\fmf{dashes}{v1,v2}
					\fmf{dashes}{o1,v2}
					\fmfv{decor.shape=hexacross, decor.size=4thick}{v2}
				\end{fmfgraph}
			}
			+ \ldots,
		\end{fmffile}
	\end{equation}
	where the asterisk stands for the impurity concentration $f$, the dashed line represents the scattering potential, and the solid line, the bare Green's function $g^{+}$.
	
	We should note that Eqs. \ref{eq:dyson} and \ref{eq:T_lowconc} form a non-self-consistent set of equations due to the fact that the expression for the $T$-matrix is approximated to be a functional of the bare Green's function. It is, in fact, possible to go beyond this approximation by replacing in Eq. \ref{eq:T_lowconc} the bare Green's function with the dressed one, which results in a set of equations coupling $\langle g^{+} \rangle_{\text{imp}}$ and $T$. But these so-called self-consistent approximations are beyond the scope of this work. It is also noteworthy that all diagrams with crossing interaction lines are omitted in the above approximation. This omission is safe when the concentration of defects is low and when there is no correlation between two defect sites.
	
	In this work, the $T$-matrix, hence $\Sigma$, is calculated \textit{ab initio}; see Section \ref{sec:comp_methods}. From Eq. \ref{eq:dyson}, one can calculate the new poles, \ie the renormalized squared angular frequencies, by self-consistently solving
	\begin{equation}
		\Omega^{2}_{\lambda} = \omega^{2}_{\lambda} + \Re\Sigma_{\lambda}\left[\Omega^{2}_{\lambda}\right]
	\end{equation}
	and the phonon-isotope scattering rates in the self-energy relaxation-time approximation is given by
	\begin{equation}\label{eq:Wphiso}
		W^{\text{ph-iso}}_{\lambda} = -\omega^{-1}_{\lambda}\Im\Sigma_{\lambda}.
	\end{equation}
	Specifically, we consider the following three low-concentration approximations -- lowest order (0B), 1$^\text{st}$ Born (1B), and full Born (FB) approximation -- for the $T$-matrix. These are given by
	\begin{align}
		T^{\text{0B}} &= fV, \\
		T^{\text{1B}} &= f\left(V + Vg^{+}V\right), \text{ and} \\
		T^{\text{FB}} &= f\left(1 - Vg^{+}\right)^{-1}V.
	\end{align}
	They correspond to the retention of the terms on the right hand side of Eq. \ref{eq:T_lowconc} up to the first, second, and infinite order, respectively.
	
	In this work, we sum the phonon-isotope scattering rates with the 3- and 4-phonon scattering rates using the Matthiessen rule for computing the phonon transport. This rule essentially states that the impurity scattering channel is independent of the rest. Note that Eq. \ref{eq:Wphiso} considers only the out-scattering of phonons. We will ignore the in-scattering for this scattering channel which, in general, is expected to weaken the scattering rates somewhat. This approximation is expected to hold in the high-temperature regime where the phonon-isotope scattering does not entirely dominate the dissipation of the phonon current. In this work, we will neglect the phonon energy renormalization effects as they are small. Furthermore, we will use the on-shell approximation for the scattering rates, \ie take $\Im\Sigma_{\lambda} = \Im\Sigma(\omega_\lambda)$ in Eq. \ref{eq:Wphiso}.
	
	\subsection{The Tamura model and the virtual crystal approximation (VCA)}
	
	Following Tamura's original work \cite{tamura_isotope_1983}, the hamiltonian of the ionic system can be formally written as
	\begin{equation}
		H = H_{0} + H_{\text{I}}.
	\end{equation}
	The first term is
	\begin{equation}
		H_{0} = \dfrac{1}{2}\sum_{\alpha l \tau} M_{\tau} \dot{u}_{\alpha l \tau}^{2} + U
	\end{equation}
	with $l$ labeling the unit cell in the crystal. $\mathbf{u}$ is the displacement of atom $\tau$ from the equilibrium position, and $U$ is the harmonic interatomic potential. The second term is
	\begin{equation}
		H_{\text{I}} = \dfrac{1}{2}\sum_{\alpha l \tau} \left( M'_{l\tau} - M_{\tau} \right) \dot{u}_{\alpha l \tau}^{2},
	\end{equation}
	where $M'_{l\tau}$ is an isotope mass at basis site $\tau$ in unit cell $l$.
	
	Note that the above separation of the hamiltonian is formally correct for any value of the basis atom mass $M_{\tau}$. That is, $M_{\tau}$ acts as a positive, continuous gauge that can be arbitrarily chosen. If $M_{\tau}$ is chosen to be too far from  the actual isotopic masses of $\tau$, then the phonon energies will be extremely different from the measured values, and a massive correction has to be sought from scattering theory. In Ref. \cite{maris1966vibrational}, $M_{\tau}$ was set to the heaviest-isotope mass. This can be a reasonable choice since the isotope masses of any atom are very similar and the phonon spectrum is close to the real-life one. Tamura, on the other hand, set $M_{\tau}$ to the (VCA) average mass of all the isotopes. This, again, seems like a reasonable choice for this gauge, because the calculated phonon energies are also close to reality. Nevertheless, we take a closer look at this latter choice.
	
	The VCA mass is given by:
	\begin{equation}
		M^{\text{VCA}}_{\tau} = \sum_{i}{f^{i}_{\tau}}M^{i}_{\tau},
	\end{equation}
	where $f^{i}$ is the natural fractional abundance of the $i$-th isotope of the considered atom. As $\sum_{i}f^{i} = 1$, $T^{\text{0B}}$ trivially vanishes in the VCA. As such, both the first non-zero lineshift and linewidth come from the $T^{\text{1B}}$ diagram. This VCA-1B approximation is called the Tamura model of the phonon-isotope scattering \cite{tamura_isotope_1983}. The scattering rate in this approximation is given by
	\begin{align}\label{eq:Wphiso_Tamura}
		W_{\text{Tamura},s\mathbf{q}} =& \dfrac{\pi\omega^{2}_{s\mathbf{q}}}{2N_{\mathbf{q}}} \sum_{s'\mathbf{q}'}\sum_{\tau}\sum_{d_{\tau} \in \text{all}}f_{d_{\tau}}\left(\dfrac{\Delta M^{(d_{\tau})}_{\tau\tau}}{M_{\tau}}\right)_{\text{VCA}}^{2} \nonumber \\ 
		&\times\left|\sum_{\alpha}\phi^{\tau\alpha}_{s}(\mathbf{q})\phi^{\tau\alpha}_{s'}(\mathbf{q}')\right|^{2}\delta(\omega_{s\mathbf{q}} - \omega_{s'\mathbf{q}'}),
	\end{align}
	where $N_{\mathbf{q}}$ is the number of unit cells in the crystal, and the sum over $\tau$ is restricted to the primitive cell. 
	
	To systematically improve over the Tamura model, the following consequences of the VCA should be considered:
	
	1) The Tamura picture has the peculiar feature that $100\%$ of the atoms may host a defect, as illustrated in the top panel of Fig. \ref{fig:vca_vs_dib}. This is because every type of isotope is a substitutional defect at the corresponding VCA atom species site, and their fractional abundances sum to unity. This, in turn, indicates that the infinite-order summation of all diagrams linear in the impurity concentration, as given in Eq. \ref{eq:T_lowconc}, could potentially be inadequate. On the other hand, this choice also ensures that the scattering potential of the high-abundance isotope will be weak compared to the one of the low-abundance isotope. As such, higher order terms in $f$ in the Born series are small, as has been shown by Tamura \cite{tamura_isotope_1983}. So it might still be fine to carry out a full Born approximation on top of the VCA regardless of the fact that there is always at least one defect that has a very high concentration. We will show later that this strategy of minimizing the scattering potential does not guarantee that the 1B scattering theory suffices, and the Tamura model may fail on account of being a low-order perturbation theory. 
	
	2) Another issue is related to the fact that the impurity-averaging procedure to calculate $\langle g^{+} \rangle$ is performed assuming a random distribution of defects \cite{mahan2000many}. In a system where every atomic site hosts a substitution, the spatial randomization of defects can often lose its meaning. This is easy to see for any isotopic substitution for which $f \lessapprox 1$. It is not clear how one can overcome this formal issue.
	
	3) Finally, the perturbation due to an isotopic substitution at the \textit{fictitious} VCA atom may be significantly smaller than that of the same substitution at the \textit{real} atom, which we argue is the true scattering potential. This issue has implications for both the 1B and FB theories. We shall show later that it can have important ramifications for the phonon thermal transport if there is significant isotopic variation of the heavier of the basis atoms.
	
	As a simple remedy for these pathologies and to satisfy the correspondence principle discussed earlier, we propose to use the dominant isotopes instead of the VCA to construct the phonon ground state. This will be described in the next section.
	
	\subsection{Dominant-isotope background method}
	
	\begin{figure}[h!]
		\includegraphics[width=7cm]{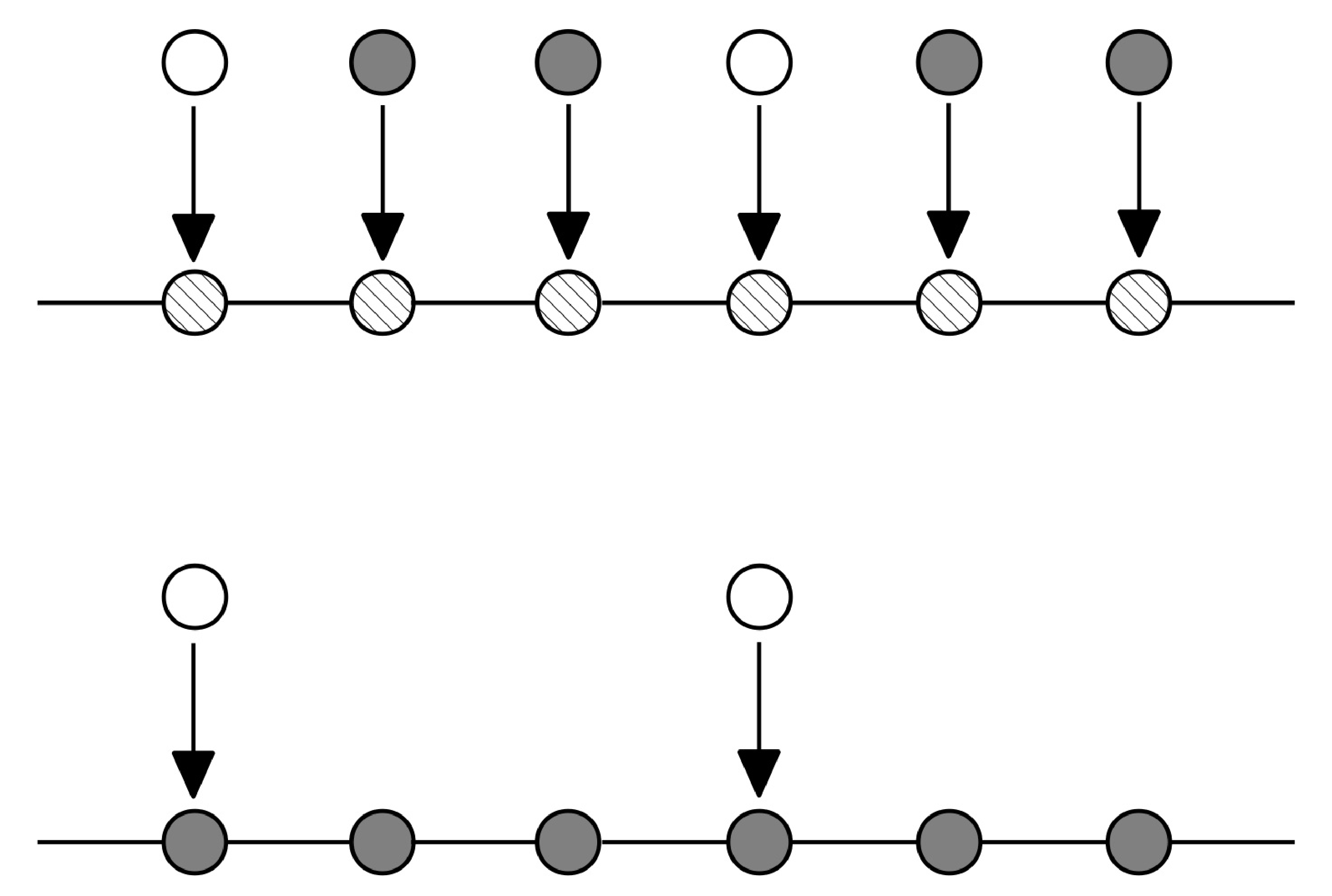}
		\caption{Cartoon illustrating the difference between the Tamura (top) and the DIB (bottom) approaches to the isotope substitution problem in a 1D monatomic chain. The hatched circles represent VCA atoms created from the average of the dominant (solid circles) and the minority (empty circles) isotopes. Arrows symbolize substitution.}
		\label{fig:vca_vs_dib}
	\end{figure}
	
	In the {\it dominant-isotope background} (DIB) method, proposed here, the phonon ground state is constructed entirely from the dominant isotope of every basis atom, as illustrated in the bottom panel of Fig. \ref{fig:vca_vs_dib}. The minority abundance isotopes are then treated as substitutional defects. As such, by construction, this is a low-concentration formulation of the isotope-scattering problem. In this case, the use of the resummation in Eq. \ref{eq:T_lowconc} is a good approximation, since the terms linear in $f$ are guaranteed to dominate over the higher-power ones. We recall that this approach is identical to the standard way substitutional defects are treated in the literature, for example in Refs. \cite{nebil_2014, katre2017exceptionally, fava2021dopants}.
	
	In contrast to the VCA case, in the DIB method, the $T^{\text{0B}}$ diagram, provides a small, frequency-dependent energy renormalization. There is no linewidth from this term, since the scattering potential is purely real. Both the $T^{\text{1B}}$ and $T^{\text{FB}}$ approximations give non-zero line shifts and line widths. In particular, the phonon-isotope scattering rate in the DIB-1B approximation is 
	\begin{align}\label{eq:Wphiso_DIB1B}
		W_{\text{DIB-1B},s\mathbf{q}} =& \dfrac{\pi\omega^{2}_{s\mathbf{q}}}{2N_{\mathbf{q}}} \sum_{s'\mathbf{q}'}\sum_{\tau}\sum_{d_{\tau} \in \text{minority}}f_{d_{\tau}}\left(\dfrac{\Delta M^{(d_{\tau})}_{\tau\tau}}{M_{\tau}}\right)_{\text{DIB}}^{2} \nonumber \\ 
		&\times\left|\sum_{\alpha}\phi^{\tau\alpha}_{s}(\mathbf{q})\phi^{\tau\alpha}_{s'}(\mathbf{q}')\right|^{2}\delta(\omega_{s\mathbf{q}} - \omega_{s'\mathbf{q}'}),
	\end{align}
	where $M_{\tau}$ is now the mass of the dominant isotope of atom $\tau$ in the primitive cell. Note especially that the summation over defects runs only through the minority isotopes. This expression should be compared with the Tamura expression, Eq. \ref{eq:Wphiso_Tamura}. The derivation of both Eqs. \ref{eq:Wphiso_Tamura} and \ref{eq:Wphiso_DIB1B} can be found in Appendix A.
	
	Existing \textit{ab initio} phonon-transport codes can readily offer Eq. \ref{eq:Wphiso_DIB1B} as an option along with Eq. \ref{eq:Wphiso_Tamura} for the users to choose from. This is already done in the \texttt{elphbolt} code \cite{protik2022elphbolt} version 1.1. 
	
	It is interesting to consider a maximally impure case, for example, where an atom has two isotopes of equal abundance. In this case, the DIB theory is trivially discontinuous, since the ratio of the masses is not equal to its inverse. In this particular scenario, the VCA method is, however, continuous. The discontinuity in the DIB theory though is natural, since the answer to the question of which isotope constitutes the pure system is ambiguous. For this case, and other similar cases, the supercell phonon unfolding method \cite{thebaud2020success} is better suited.
	
	In this work, we go beyond the Tamura model (\ie VCA-1B), in three ways: (i) by applying the full Born approximation while retaining the VCA masses (VCA-FB); (ii) by using the dominant-isotope background masses instead of the VCA masses but retaining the \first Born approximation (DIB-1B); and (iii) by using the dominant-isotope background masses along with the full Born approximation (DIB-FB).
	
	\section{Computational methods}\label{sec:comp_methods}
	
	We compute the $2^{\text{nd}}$-order interatomic force constants for the considered materials -- silicon, lithium fluoride, boron arsenide, and gallium nitride -- from density-functional perturbation theory using \texttt{Quantum Espresso} \cite{giannozzi2009quantum, giannozzi2017advanced}. These are fed into \texttt{elphbolt} \cite{protik2022elphbolt}, which calculates the necessary phonon ground-state properties. The phonon-isotope scattering $T$-matrix formalism has also been implemented recently in the latter. The bare phonon Green's function is calculated using the analytical tetrahedron method \cite{lambin1984computation}, fixing the typographical errors in the expressions for the real part of the resolvent following Ref. \cite{eyert2012augmented}.
	
	For the case of the wurtzite GaN, the phonon thermal conductivity is also calculated. For this, we generate the necessary $3^{\text{rd}}$-order interatomic force constants using \texttt{thirdorder} \cite{li2014shengbte}. The 4-phonon scattering rates are computed using the codes \texttt{FourPhonon} and \texttt{fourthorder} \cite{han2022fourphonon}. The results are then passed to \texttt{elphbolt} to iteratively solve the phonon Boltzmann transport equation to calculate the phonon thermal conductivity. This workflow has previously been detailed in Ref. \cite{elhajhasan2023joined}.
	
	Details regarding the various computational parameters are given in Appendix B.
	
	\section{Numerical results and discussion}
	
	\begin{table*}[htb]
		\begin{tabular}{ |c|c|c|c|c|c|c|c|}
			\hline
			& \multicolumn{3}{|c|}{VCA} & \multicolumn{3}{|c|}{DIB} & \\
			\hline
			Atom & $M$ & $-\Delta M/M$ & $f$ & $M$ & $-\Delta M/M$ & $f$ & $r$ \\
			\hline
			Si   & $28.0855$  & $3.87\times 10^{-3}$ & $0.9223$  &  $27.9769$ & $0$ & $1$ & $1.083$ \\
			&           & $-3.17\times 10^{-2}$ & $0.0467$ &   & $-3.57\times 10^{-2}$ & $0.0467$ &\\
			&           & $-6.72\times 10^{-2}$ & $0.031$  &   & $-7.14\times 10^{-2}$ & $0.031$ & \\
			\hline
			Ga   & $69.7232$  & $1.14\times 10^{-2}$ & $0.601$  &  $68.9256$ & $0$ & $1$ & $1.703$ \\
			&           & $-1.72\times 10^{-2}$ & $0.399$  &   & $-2.90\times 10^{-2}$ & $0.399$ &\\
			\hline
			N   & $14.0068$  & $2.63\times 10^{-4}$ & $0.9963$  &  $14.0031$ & $0$ & $1$ & $1.004$ \\
			&           & $-7.09\times 10^{-2}$ & $0.0037$  &   & $-7.12\times 10^{-2}$ & $0.0037$ &\\
			\hline
			Li   & $6.9417$  & $-1.07\times 10^{-2}$ & $0.9258$  &  $7.0160$ & $0$ & $1$ & $1.057$\\
			&           & $1.33\times 10^{-1} $ & $0.0742$  &   & $1.43\times 10^{-1} $ & $0.0742$ &\\
			\hline
			F   & $18.9984$  & $0$ & $1$  &  $18.9984$ & $0$ & $1$ & $-$\\
			\hline
			B   & $10.8120$  & $-1.82\times 10^{-2}$ & $0.8020$  &  $11.0093$ & $0$ & $1$ & $1.203$ \\
			&           & $7.39\times 10^{-2}$ & $0.1980$  &   & $9.05\times 10^{-2}$ & $0.1980$ &\\
			\hline
			As   & $74.9216$  & $0$ & $1$  &  $74.9216$ & $0$ & $1$ & $-$ \\
			\hline
		\end{tabular}
		\caption{Comparison of on-site mass defects due to the natural isotopic mix in the VCA and DIB models. Columns 2 and 5 show the corresponding host atom masses where the substitutions happen, respectively.}
		\label{tab:defect}
	\end{table*}
	
	In Table \ref{tab:defect}, we compare the mass perturbations, Eq. \ref{eq:massdiff}, in the VCA and DIB models due to the isotope substitutions for the set of considered materials. The last columns shows the following ratio:
	\begin{equation}\label{eq:1B-strength-ratio}
		r = \frac{\sum_{d_{\tau \in \text{minority}}} f_{d_{\tau}}\left( \Delta M^{(d_{\tau})}_{\tau\tau}/M^{\text{DIB}}_{\tau} \right)^{2}} {\sum_{d_{\tau \in \text{all}}} f_{d_{\tau}}\left( \Delta M^{(d_{\tau})}_{\tau\tau}/M^{\text{VCA}}_{\tau} \right)^{2}}.
	\end{equation}
	It gives a measure of the relative strength of the phonon-isotope scattering between the DIB and VCA approaches in the 1B approximation. In all cases, the DIB-1B scattering is stronger than the VCA-1B scattering to a varying degree that depends on how far the VCA mass is from the dominant isotope's mass. It should be noted, however, that the effective interaction in the FB approximation may be weaker or stronger than in the 1B approximation. In what follows, we study the various theories of the phonon-isotope interaction in a set of selected materials.
	
	\subsection{Silicon}
	
	\begin{figure}[h!]
		\includegraphics[width=9.5cm]{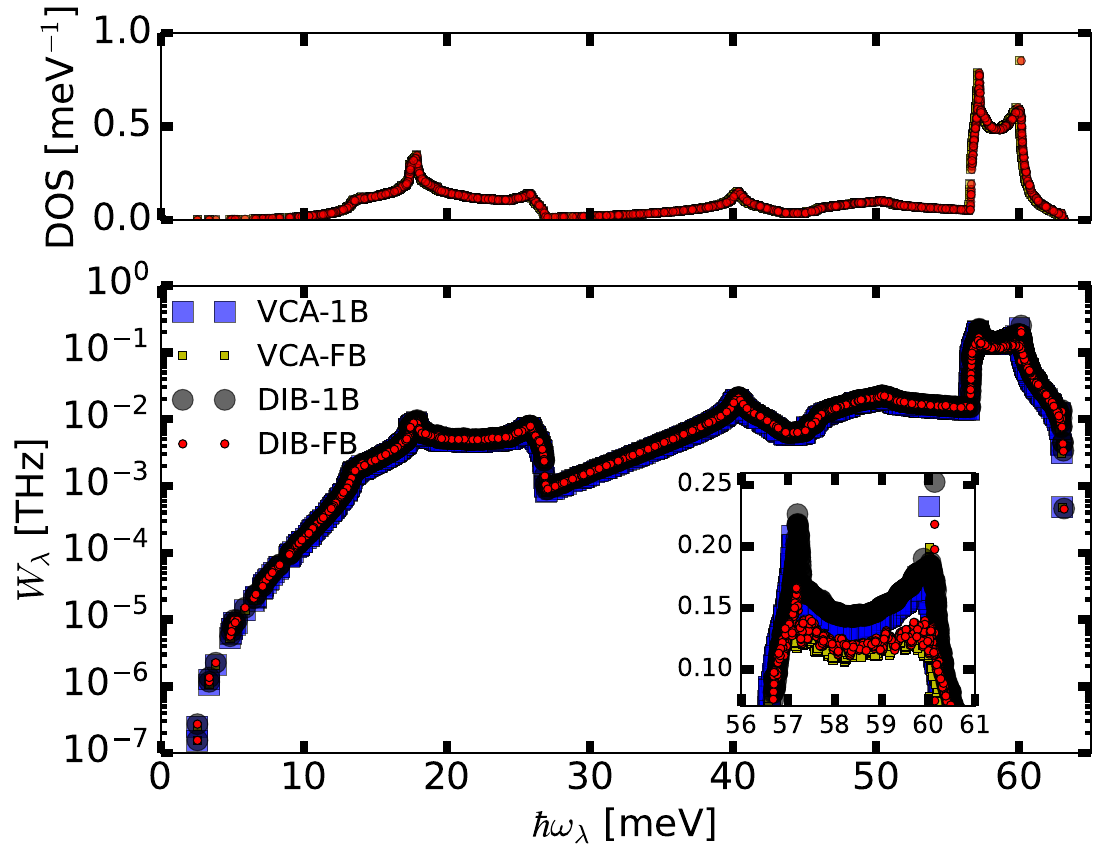}
		\caption{Phonon density of states (top panel; yellow squares for VCA and red dots for DIB) and phonon-isotope scattering rates (bottom panel; 4 different models) for Si.}
		\label{fig:Si_Wphiso}
	\end{figure}
	
	The phonon-isotope scattering rates along with the phonon density of states (DOS) of Si are shown in Fig. \ref{fig:Si_Wphiso}. The scattering rates for the acoustic phonons are essentially the same in the Tamura model (VCA-1B) and the models beyond. The negligible difference between the VCA and DIB results is due to the fact that the corresponding masses are nearly the same, as shown in Table \ref{tab:defect}. The ratio $r$ is therefore nearly unity. In the range of transverse optical (TO) phonons, \ie $57-60$~meV (see inset), there exists, however, a non-negligible difference between the \first Born (VCA-1B, DIB-1B) and the full-Born (VCA-FB, DIB-FB) theories, the former giving up to $\times 1.4$ stronger scattering over the latter. This difference, however, is not going to have any discernible effect on the phonon transport since these TO bands are flat. The difference appears where the phonon DOS is large. This is expected since in the Born series, increasingly higher powers of the Green's function appear, and the imaginary part of the Green's function is proportional to the delta function. This demonstrates a failure of the Tamura model for the optical phonons due to the premature truncation of the Born series.

	\subsection{Lithium Fluoride}
	\begin{figure}[h!]
		\includegraphics[width=9.5cm]{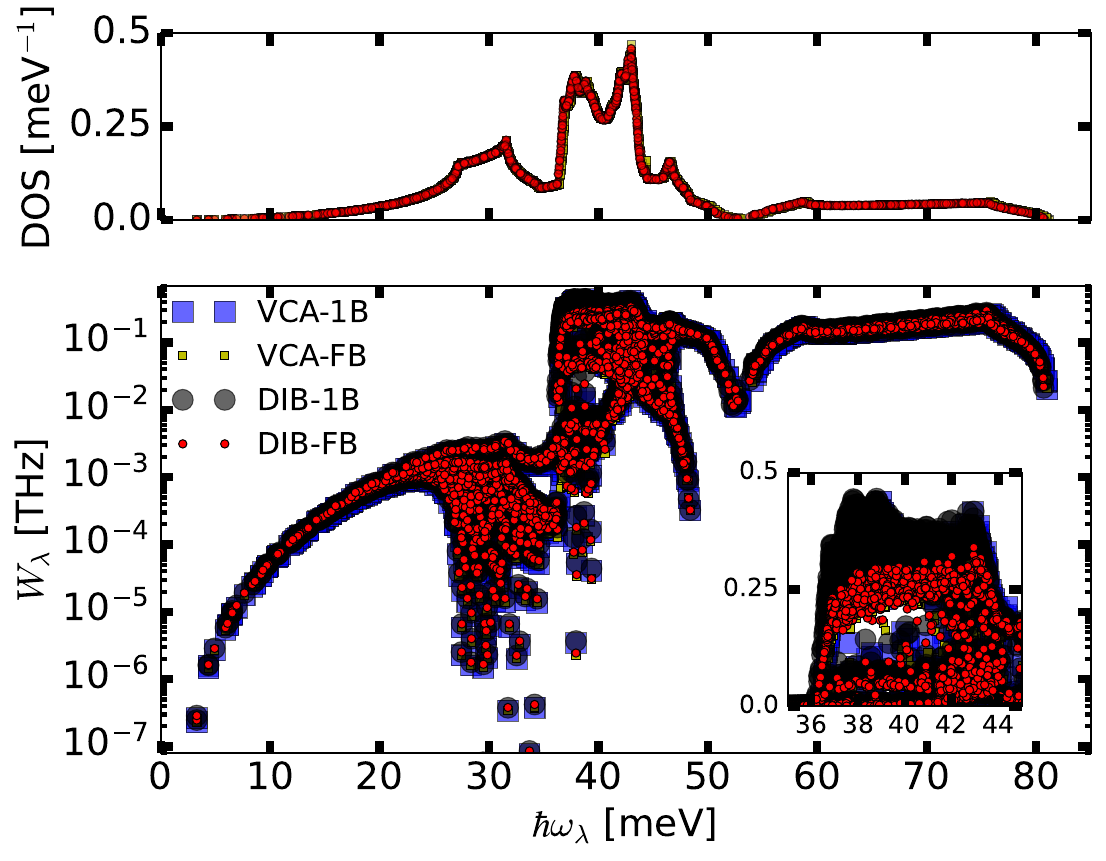}
		\caption{Same as Fig. \ref{fig:Si_Wphiso} but for LiF. Note that in the lower panel inset, the blue (yellow) symbols are hardly visible, since they mostly overlap with the black (red) symbols.}
		\label{fig:LiF_Wphiso}
	\end{figure}
	
	The results for LiF are given in Fig. \ref{fig:LiF_Wphiso}. First, we note from Table \ref{tab:defect} that the heavier F atom that dominates the acoustic vibrational modes is isotopically pure, while the VCA and DIB masses of the lighter Li atom are nearly equal. As a consequence, just like in the Si case, the acoustic phonon-isotope scattering rates agree among all four theories. Again, we see a significantly larger (up to $\times 1.5$) difference between the 1B and FB theories for the weakly-dispersive TO phonons in the $37-43$ meV range, where the DOS is large. The reason for the observed difference is identical to that for Si above. This again demonstrates a shortcomings of the Tamura model for the optical phonons due to the premature truncation of the Born series.
	
	\subsection{Boron Arsenide}
	\begin{figure}[h!]
		\includegraphics[width=9.5cm]{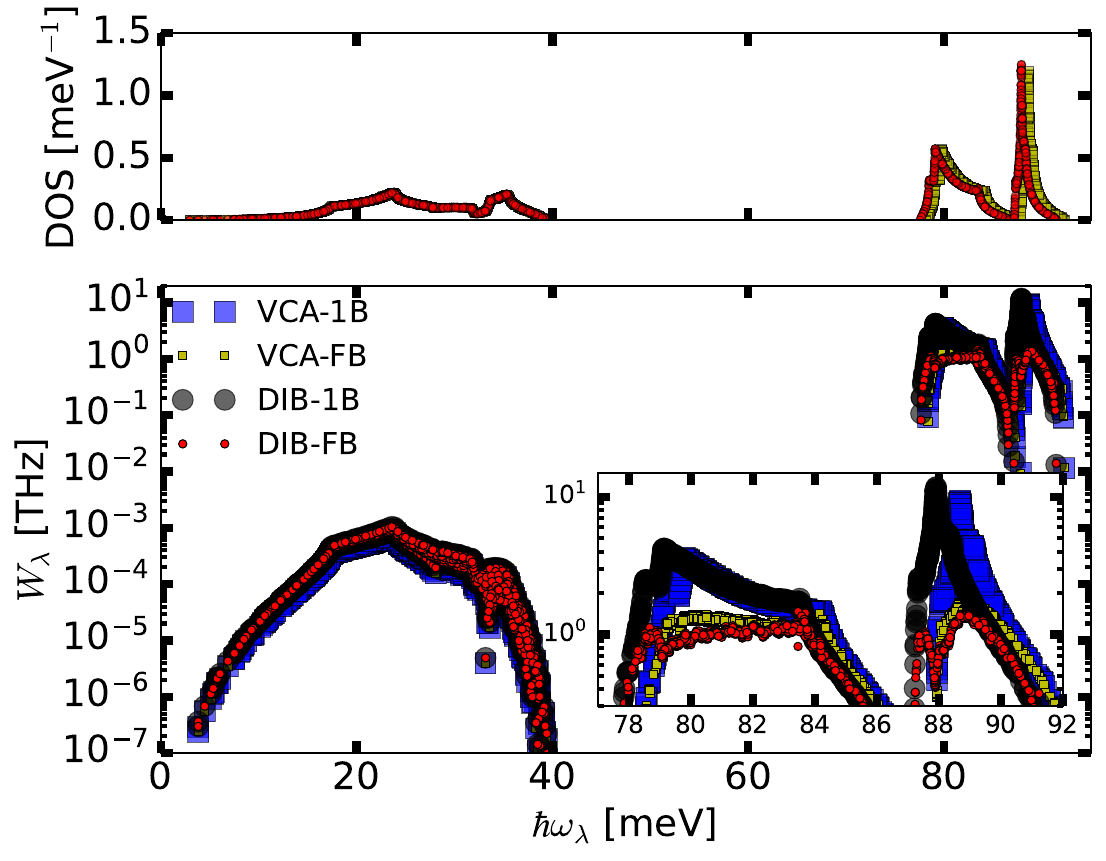}
		\caption{Same as Fig. \ref{fig:Si_Wphiso} but for BAs.} 
		\label{fig:BAs_Wphiso}
	\end{figure}
	
	In cubic BAs, the heavier As atom is isotopically pure, while the B atom has two isotopes, $80.2\%$ $^{10}$B and $19.8\%$ $^{11}$B, as shown in Table \ref{tab:defect}. Since the $r$ value for B is $1.2$, we can expect a non-negligible difference between the VCA and DIB calculations. The DOS and isotope scattering rates are shown in Fig. \ref{fig:BAs_Wphiso}. For the acoustic phonons, both DIB scattering rates are about a factor $r$ larger than the VCA ones, with no difference coming from the choice of the 1B or FB model. The acoustic phonon-isotope scattering rates in this material are weak because the heavy As atom that dominates acoustic vibrations is isotopically pure. Now, while the acoustic phonons are the dominant carriers of heat, because the acoustic phonon-isotope scattering rates here are significantly weaker than the phonon-phonon counterparts \cite{ravichandran2020phonon}, the significant difference between the VCA and DIB theories does not lead to a proportional difference in the phonon transport properties. 
	
	For the optical phonons that are dominated by the motion of the light B atom, the situation is slightly more complex. First, their VCA energies are shifted higher compared to the DIB ones as can be seen in the DOS. This alone is not expected to cause a large change in the phonon transport, since the shift is less than 1 meV. The FB scattering rates are found to be up to 10 times smaller than the 1B ones (note the logarithmic scale). Large differences again occur where the DOS is high, as has been seen earlier in both Si and LiF. These flat optical phonons have been shown to contribute negligibly compared to the acoustic phonons in BAs \cite{ravichandran2020phonon}, so this large difference between FB and 1B will not directly impact the phonon thermal conductivity within the phonon Boltzmann transport formalism. 
	
	Since the optical phonon-isotope scattering rates are so large in this material, it is worth checking whether the quasiparticle picture is compromised for these phonons. To this end, we apply the Landau criterion which checks whether the scattering rate of a phonon mode is significantly smaller than the frequency of that mode: $W_{s\mathbf{q}}^{\text{ph-iso}} << \omega_{s\mathbf{q}}$ \cite{SmithHenrik1989Tp/H}. We find that for the most strongly damped phonons around $88$ meV ($\approx 21$ THz), $W^{\text{ph-iso}}_{\text{1B}}$ is larger than $10$ THz (black and blue symbols), while $W^{\text{ph-iso}}_{\text{FB}}$ is close to $1$ THz (red and yellow symbols). That is, the FB theory preserves the quasiparticle picture by a safe margin and consequently allows subsequent application of methodologies like the Boltzmann transport equation. The same, however, cannot be said for the 1B theories. For this material, too, we see the need to go beyond the Tamura model on account of the failure of the low-order perturbation theory.
	
	\subsection{Gallium Nitride}
	\begin{figure}[h!]
		\includegraphics[width=9.5cm]{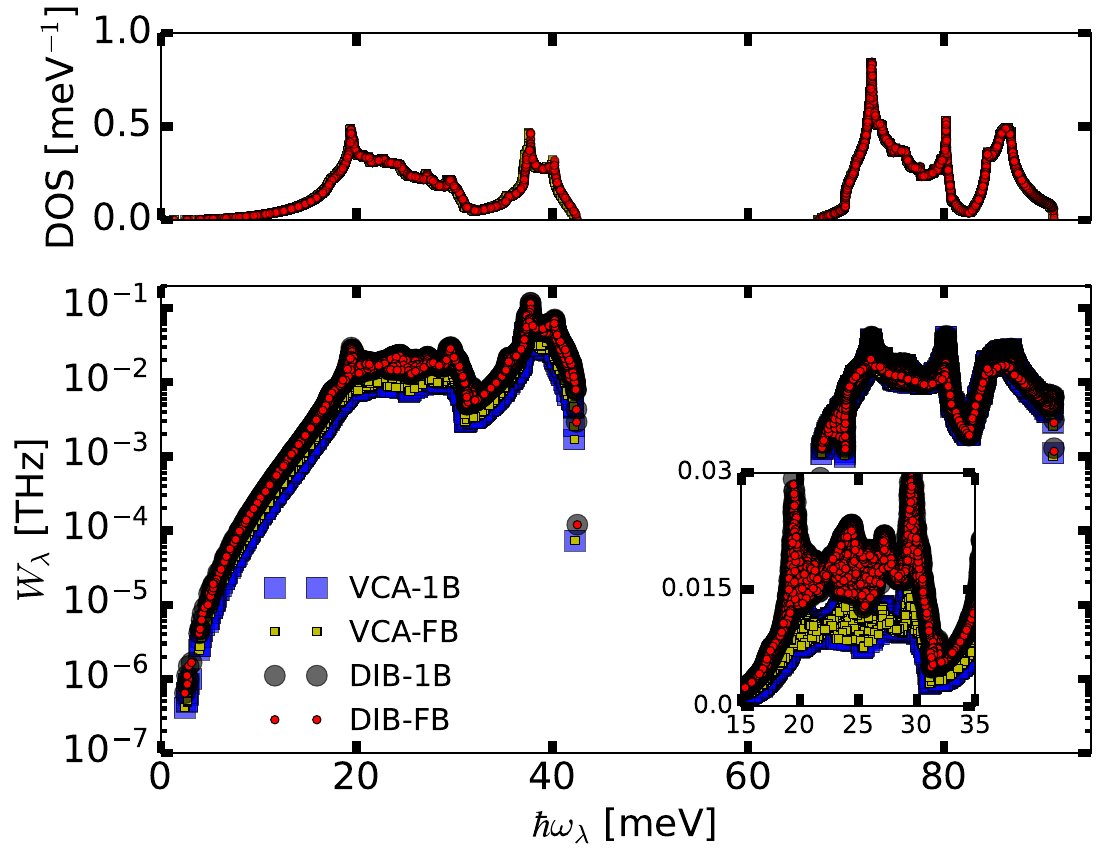}
		\caption{Same as Fig. \ref{fig:Si_Wphiso} but for wurtzite GaN.}
		\label{fig:GaN_Wphiso}
	\end{figure}
	
	For wurtzite GaN, we find from Table \ref{tab:defect} that for the heavier Ga atom that dominates the acoustic vibrational modes, the DIB perturbation is significantly larger than the VCA counterpart, with an $r$ factor of $1.7$. In other words, the true substitutional defect-scattering potential at the Ga site is significantly stronger than that given by the VCA theory. As a result, in the entire acoustic and low-lying optical phonon region ($0-40$ meV), the phonon-isotope scattering rates in the DIB model are by a factor $r$ stronger compared to the VCA ones, as can be seen in Fig. \ref{fig:GaN_Wphiso}. The inset zooms in on the $15-35$ meV region, which has previously been shown to give a large phonon thermal conductivity contribution when both the 3- and 4-phonon scattering are considered \cite{elhajhasan2023joined}. As such, one can expect a noticeable reduction of the thermal conductivity in the DIB theory compared to the VCA. We also note that for all phonons below the gap, no improvement is achieved by going beyond the 1B theory for either the VCA or the DIB methods. This, in turn, indicates that for these phonons, the true scattering potential is simply unreachable in the VCA model, even by improving the scattering theory by going from the 1B to FB case.
	
	\begin{figure}[ht]
		\includegraphics[width=9.5cm]{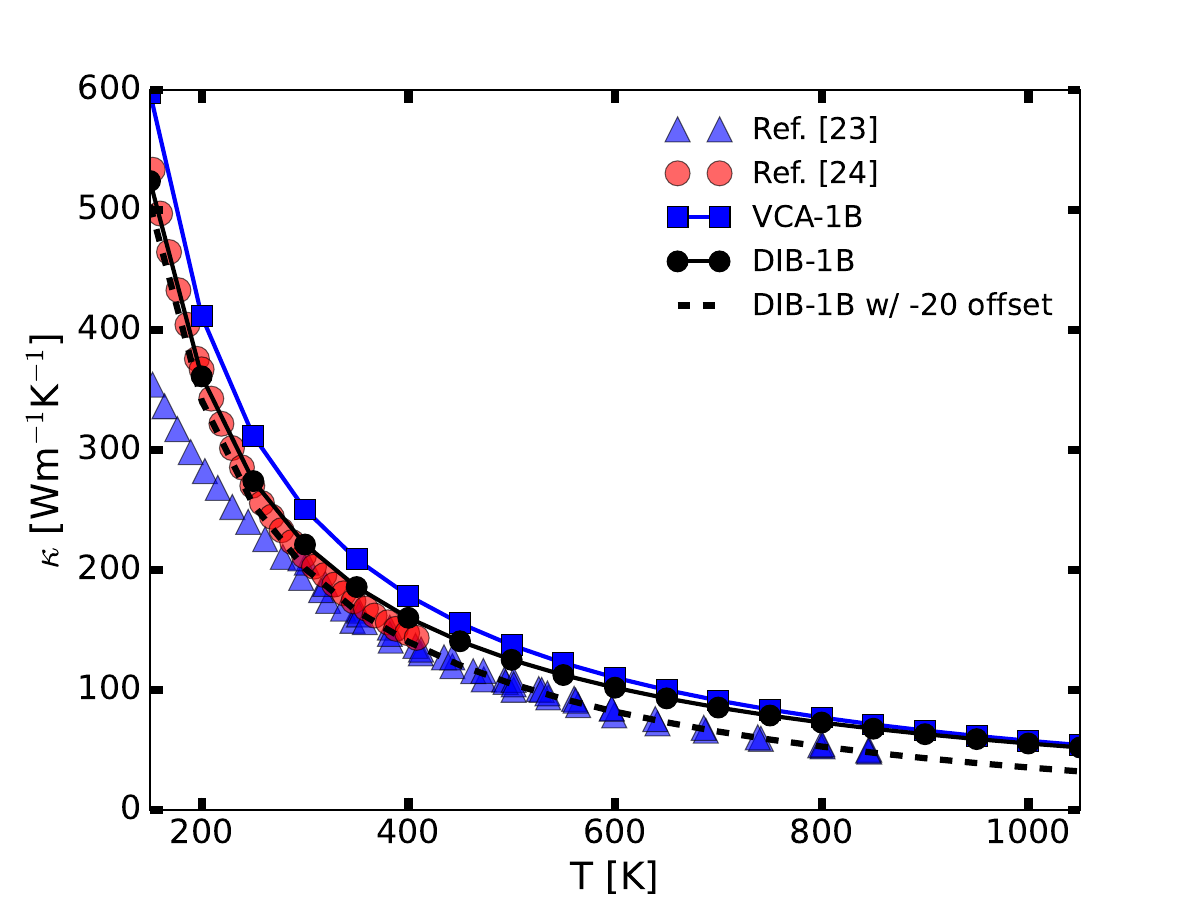}
		\caption{Isotropic average of the phonon thermal conductivity, $\kappa$, as a function of temperature for wurtzite GaN. For comparison, experimental data from Ref. \cite{zheng2019thermal} (blue triangles) and Ref.  \cite{inyushkin2020high} are shown (red circles). To highlight that the DIB-1B shows excellent agreement with the experimentally observed temperature dependence, we repeat this result with a $-20$~Wm$^{-1}$K$^{-1}$ offset (dashed line).}
		\label{fig:GaN_kappa_T}
	\end{figure}
	
	For the high-lying optical phonons, on the other hand, the situation is similar to what we have seen for the other materials studied thus far. In the $70-80$ meV range, for example, both the 1B theories predict a two times stronger scattering compared to the FB theories. This is, however, expected to have a minor impact on the phonon thermal conductivity, since the spectral contribution for this quantity in that energy range has been shown to be small \cite{elhajhasan2023joined}.
	
	In order to quantify how the DIB vs VCA theories affect the phonon thermal conductivity we calculate its temperature dependence. We include only the 1B theories in these calculations, since very little difference was observed between the 1B and FB isotope scattering rates for the heat-carrying acoustic phonons. We use the 4-phonon scattering rates calculated with the \texttt{FourPhonon} code on a coarse mesh using the VCA masses for both the VCA and DIB sets of calculations in \texttt{elphbolt}. This is fully justified, since both theories produce essentially the same ground-state energies, as can be seen from the DOS in Fig. \ref{fig:GaN_Wphiso}. red Since the anisotropy of the thermal conducitivity is small, cross-plane to in-plane ratio $\approx$ $1.04$ and $1.02$ at $150$ K and $1050$ K, respectively, we consider only the isotropic average of this quantity. In Fig. \ref{fig:GaN_kappa_T}, we plot the thermal conductivity along with measurements from Ref. \cite{zheng2019thermal} (blue triangles) and from Ref. \cite{inyushkin2020high} (red circles). We find that at high temperatures close to $1000$ K, the Tamura (blue square-line) and DIB-1B (black circle-line) theories converge. This is the expected behavior, because the phonon-isotope scattering rates are temperature independent, whereas the 3- and 4-phonon scattering rates increase with increasing temperature. The absolute difference between the Tamura and DIB-1B thermal conductivities increases with decreasing temperature, with the latter theory always predicting a lower value. At $150$ K, the DIB-1B theory predicts a lower thermal conductivity by about $74$ Wm$^{-1}$K$^{-1}$ compared to the Tamura model. The DIB-1B prediction is closer to the measurements in the entire temperature range considered here.
	
	It is interesting to observe that the DIB-1B theory shows excellent agreement with the measured temperature dependence in the $> 300$ K range. To highlight this, we plot the curve again with a constant vertical offset (dashed black line). In contrast, it is not possible to get the Tamura results to agree with the experimental data in the high temperature regime by a similar constant shift. This shows that the commonly held idea that the temperature dependence of the thermal conductivity around room temperature and above is entirely determined by phonon-phonon scattering is generally an oversimplification. Earlier calculations excluding the 4-phonon scattering also led to a weaker temperature dependence than measured \cite{inyushkin2020high,zheng2019thermal,lindsay2012thermal,lindsay2013phonon}. In the present work, by including both the 3- and 4-phonon scattering and using the DIB-1B theory, we find good agreement with the measured temperature dependence of this transport coefficient. To our knowledge, the temperature-dependent thermal conductivity of wurtzite GaN including both these interactions has not been reported in the literature before. The reason why 4-phonon scattering is important in this material has already been discussed in Ref. \cite{elhajhasan2023joined}.
	
	Now, the reason for the overestimated theoretical predictions in the regime above $300$~K could be partly due to the fact that the additional dissipation of the phonon current due to the phonon-electron scattering has been ignored in this theoretical modeling. This particular scattering channel is weakly temperature dependent at high temperatures \cite{ziman2001electrons}. The theoretical predictions also significantly overestimate the thermal conductivity in the $< 300$ K regime compared to the measurements of Ref. \cite{zheng2019thermal}, although excellent agreement is found with the measurements of Ref. \cite{inyushkin2020high}. The large discrepancy with one of the experimental sets could be due to the fact that at low temperatures, the phonon-dislocation and other phonon-defect interactions become progressively more important due to the weakening of the 3- and 4-phonon scattering. Also, the phonon-electron scattering increases with decreasing temperature \cite{ziman2001electrons}. None of these additional, strongly sample-quality dependent phonon-current dissipating interactions has been included in the theoretical modeling.
	
	\section{Summary and outlook}
	In this work, we have explored three ways to go beyond the Tamura model of the phonon-isotope scattering. Here we summarize the main results:
	
	(i) We have shown the importance of considering the dominant isotope mass for every atomic species rather than relying on the average mass as used in the VCA. Our DIB theory unifies the phonon-isotope scattering theory with the more general phonon-substitution scattering theory in the appropriate limit. On the contrary, the Tamura model does not satisfy this desirable correspondence principle. We have provided an alternative 1B theory expression, Eq. \ref{eq:Wphiso_DIB1B}, to the Tamura formula, which can be readily implemented in existing phonon-transport codes. For the phonon thermal conductivity of wurtzite GaN, we have also shown that the DIB-1B theory gives better agreement with recent measurements over a wide temperature range compared to the Tamura model. This is true for both the thermal conductivity and its temperature dependence. Moreover, we have demonstrated that in this material both the 3- and 4-phonon scattering must be included in the phonon BTE to obtain good agreement with thermal-conductivity measurements at high temperatures. 
	
	(ii) For all materials studied here, we have shown that the non-perturbative FB theory gives significant corrections over the perturbative 1B approximation for the optical phonons. This indicates that a systematic improvement over the Tamura model can be achieved by using the FB formalism in the scattering $T$-matrix. In particular, for BAs, the 1B approximation overestimates the scattering rates by up to an order of magnitude, while the FB theory preserves the quasiparticle picture of the optical phonons which is a prerequisite for the application of the BTE. 
	
	Based on our work, we suggest the use of the FB theory over the 1B theory when accurate scattering rates are required for the optic phonons. We further suggest the use of the DIB approach over the VCA approach for those cases where the $r$-factor as defined in Eq. \ref{eq:1B-strength-ratio} is large for the heavier atom. In these cases, the DIB theory gives more accurate scattering rates for the acoustic phonons which, in turn, can significantly affect the phonon thermal conductivity.
	
	We conclude with a few ideas for possible future investigations. In the current approximation, the phonon-isotope in-scattering term is ignored, and consequently, this scattering channel is maximally dissipative. Including this correction could be beneficial, especially in the intermediate temperature range between the boundary-scattering and anharmonic-scattering dominated regimes. Including the in-scattering corrections can potentially also impact the calculation of the electron-phonon drag behavior \cite{gurevich1989electron}. In particular, this may be the case for materials where the electrons transfer a significant amount of momentum to the optical phonons \cite{protik2020electron}. If the true optical phonon-isotope scattering rates are significantly lower than those predicted by the Tamura model, as we have seen for the case of BAs, then the overall drag effect may be larger, since there will be less dissipation in the coupled electron-phonon system. Finally, the $T$-matrix formalism discussed here can also be applied to the electron-point defect scattering problem. This has previously been done by Kaasbjerg \cite{kaasbjerg2020atomistic} for 2D materials using the isotropic electronic band approximation. One may improve on this methodology by considering the fully anisotropic electronic structure, adding the in-scattering corrections, and combining this with the coupled electron-phonon BTEs \cite{protik2022elphbolt}.
	
	\bigskip
	\section*{Acknowledgment}
	Work supported by the Alexander von Humboldt Foundation, Germany. NHP acknowledges stimulating discussions with Jorge O. Sofo.
	
	\section*{Appendix}
	\subsection{Derivation of 1B scattering rates}
	Defining $R_{\lambda\lambda'} \equiv \left(\omega_{\lambda}^{2} - \omega_{\lambda'}^{2} + i0^{+}\right)^{-1}$ and considering a real, diagonal potential, $V^{ij} = V^{i}\delta_{ij}$, we have in the 1B approximation for a single defect
	\begin{align*}
		T^{\text{1B}}_{\lambda\lambda} &= \sum_{ij}\phi^{*i}_{\lambda}\phi^{j}_{\lambda}\left[ V_{\lambda}^{i}\delta_{ij} + \sum_{mn}V_{\lambda}^{i}\delta_{im}V_{\lambda}^{n}\delta_{nj}\dfrac{1}{N_{\mathbf{q}}}\sum_{\lambda'}\phi^{m}_{\lambda'}\phi^{*n}_{\lambda'}R_{\lambda\lambda'} \right] \\
		&= \sum_{i}V^{i}_{\lambda}\left|\phi^{i}_{\lambda}\right|^{2} + \dfrac{1}{N_{\mathbf{q}}}\sum_{\lambda'}R_{\lambda\lambda'}\left|\sum_{i}   V^{i}_{\lambda}\phi^{*i}_{\lambda}\phi^{i}_{\lambda'} \right|^{2}.
	\end{align*}
	Using the fact that $\Im R_{\lambda\lambda'} = \pi\delta(\omega_{\lambda} - \omega_{\lambda'})/2$, putting back in the defect fractions, and summing over the different types of defects, the phonon-isotope scattering rates are
	\begin{align*}
		W^{\text{ph-iso}}_{\lambda} =\dfrac{\pi\omega_{\lambda}^{2}}{2N_{\mathbf{q}}}\sum_{\tau}\sum_{d_{\tau}}&f_{d_{\tau}}\left(\dfrac{\Delta M^{(d_{\tau})}_{\tau\tau}}{M_{\tau}}\right)^{2}\sum_{\lambda'}\left|\vec{\phi}^{*\tau}_{\lambda}\cdot\vec{\phi}^{\tau}_{\lambda'}\right|^{2} \\
		\times\delta(\omega_{\lambda} - \omega_{\lambda'}).
	\end{align*}
	On-site defects can be restricted to be in the central primitive unit cell. Then $\phi$ can be replaced with $\xi$, since the phase for the central cell is trivial.
	
	\subsection{Parameters of \textit{ab initio} calculations}
	In Table \ref{tab:abinitoparams}, we list the parameters for all the materials considered in this work. The parameters of the density functional theory (DFT) and density functional perturbation theory (DFPT) calculations for wurtzite GaN, were previously given in Ref. \cite{elhajhasan2023joined}. The input and output files of all calculations are available in NOMAD \cite{draxl2019}, \url{https://doi.org/10.17172/NOMAD/2024.01.09-1}.
	\begin{table*}[h]
		\begin{tabular}{ l|c|c|c|c}
			Parameter / Material & Si & LiF & BAs & GaN \\
			\hline
			Lattice constant(s) (nm) & 0.540 & 0.390 & 0.474 & 0.315, 0.514 \\
			Plane-wave cutoff (Ry) & 45 & 100 & 80 & 150 \\
			DFT $\mathbf{k}$-mesh & 12$\times$12$\times$12 & 8$\times$8$\times$8 & 10$\times$10$\times$10 & 6$\times$6$\times$6 \\
			DFPT $\mathbf{q}$-mesh & 6$\times$6$\times$6 & 8$\times$8$\times$8 & 6$\times$6$\times$6 & 6$\times$6$\times$6 \\
			Phonon-isotope scattering $\mathbf{q}$-mesh & 24$\times$24$\times$24 & 24$\times$24$\times$24 & 24$\times$24$\times$24 & 24$\times$24$\times$24 \\
		\end{tabular}
		\caption{Computational parameters of DFT, DFPT, and phonon-isotope scattering calculations for the materials studied in this work. For the phonon transport of wurtize GaN, a 36$\times$36$\times$36 $\mathbf{q}$-mesh was used for all scattering channels. All $\delta$-functions in \texttt{elphbolt} were evaluated using the tetrahedron method.}
		\label{tab:abinitoparams}
	\end{table*}
	
	All calculations for scattering rates and transport coefficients have been carried out with \texttt{elphbolt v1.1}. We note that, by default, \texttt{elphbolt} expects the third order interatomic force constants (IFC3s) for all phonon calculations. In this work, however, for all materials except wurtzite GaN we only consider the phonon-isotope scattering rates for which the IFC3s are not needed.
	
	\bibliography{refs}
	
\end{document}